\documentclass[twocolumn,apl,showpacs]{revtex4}
\usepackage{graphicx}

\begin{document}

\title
{Spintronic Properties of Zigzag-Edged Triangular
Graphene Flakes}

\author{H. {\c S}ahin}
\email{shasan@bilkent.edu.tr}
\affiliation{UNAM-Institute of Materials Science and Nanotechnology,
Bilkent University, 06800 Ankara, Turkey}

\author{R. T. Senger}
\email{tugrulsenger@iyte.edu.tr}
\affiliation{Department of Physics, Izmir Institute of Technology,
35430 Izmir, Turkey}

\author{S. Ciraci}
\email{ciraci@fen.bilkent.edu.tr}
\affiliation{UNAM-Institute of Materials Science and Nanotechnology,
Bilkent University, 06800 Ankara, Turkey}
\affiliation{Department of Physics, Bilkent University, 06800 Ankara,
Turkey}

\date{\today}

\begin{abstract}

We investigate quantum transport properties of triangular graphene
flakes with zigzag edges by using first principles calculations.
Triangular graphene flakes have large magnetic moments which vary
with the number of hydrogen atoms terminating its edge atoms and
scale with its size. Electronic transmission and current-voltage
characteristics of these flakes, when contacted with metallic
electrodes, reveal spin valve and remarkable rectification
features. The transition from ferromagnetic to antiferromagnetic
state under bias voltage can, however, terminate the spin
polarizing effects for specific flakes. Geometry and size dependent
transport properties of graphene flakes may be crucial for
spintronic nanodevice applications.

\end{abstract}


\maketitle

\section{INTRODUCTION}

Graphene and graphene based nanostructures are focus of intensive
research activities due to their impressive material properties
\cite{novo1,novo2,zhang,morozov,lee,bolotin} and promising application
potential \cite{ultrafast, readout, wang, kim, yan, murali} in novel
electronic devices. In particular edge-localized spin polarizations
found in graphene ribbons \cite{edge-states, cohen}, flakes
\cite{hasan-transport,ezawa2,hod2}, and at defect sites
\cite{defect1, defect2} introduce magnetic properties that can be
utilized for spintronic applications. Recent
studies have also revealed the ferromagnetic ground state of graphene
nanodots\cite{hod}, triangular
shaped graphene fragments\cite{akola,kaxiras,chemphys,Rozhkov} and
graphene domains on 2D hydrocarbons\cite{hasan-apl} and the possibility
of observing spin polarized current voltage characteristics of such
graphene flakes.

While pristine graphene provides high carrier mobility and ambipolar
behavior, semiconductor nanoscale materials having tunable bandgap are
more desirable from the perspective of potential nanoelectronics
applications. In this context, recent efforts have been devoted to
precise controlling electronic and magnetic properties of graphene
sheets by functionalization via adatom adsorption.  The synthesis of a
2D hydrocarbon in honeycomb structure, namely
graphane,\cite{elias,sofo,boukhvalov, Graphene to graphane} is one of
the succesfull example for such functionalization. Very recently, we
reported the possibility of obtaining tunable bandgap and magnetization
through dehydrogenation of domains on 2D graphane and graphane
nanoribbons.\cite{hasan-apl, graphaneNR} Stability and electronic
properties of graphene flakes uniformly functionalized by methyl
(CH$_{3}$), phenyl (C$_{6}$H$_{5}$) and nitrophenyl
(C$_{6}$H$_{4}$NO$_{2}$) groups were also discussed
earlier.\cite{Bekyarova, Boukhvalov, Pei}

Recent experimental observations\cite{Warner, girit,
Dresselhaus,dobrik,ozyilmaz} and theoretical studies
\cite{gan,olcay,wei} show that the electronic, magnetic and conductance properties of graphitic fragments can be changed significantly upon the termination of their edges. In addition to purely zigzag and armchair edged graphene, experimental verification of the existence of alternating series of zigzag and armchair segments at the edges and energetics of reconstructions have
also been reported.\cite{hakinen} Originating from the antiferromagnetic ground state of zigzag edges, adatom\cite{hasan-transport} and topology\cite{kuc}
dependent trends in electronic properties of rectangular flakes have also investigated.

In this work, we study graphene flakes having equilateral
triangular shapes with zigzag edges ($n$-TGF), where $n$ denotes
the number of edge hexagonal cells in one side of the triangle.
The flakes have been considered as bare (C$_{n^2+4n+1}$), each
edge atom being saturated with one (C$_{n^2+4n+1}$H$_{3n+3}$) or
two hydrogen atoms (C$_{n^2+4n+1}$H$_{6n+6}$). We find that these
flakes have large spin magnetic moment values of $4(n-1)$,
$(n-1)$, and $2(n-1)$, respectively, in units of $\mu_B$. When
these TGFs have been contacted with thin metallic electrodes we
calculate that the current running through them gets both
spin-polarized and rectified.

\begin{figure}
\includegraphics[width=8.5cm]{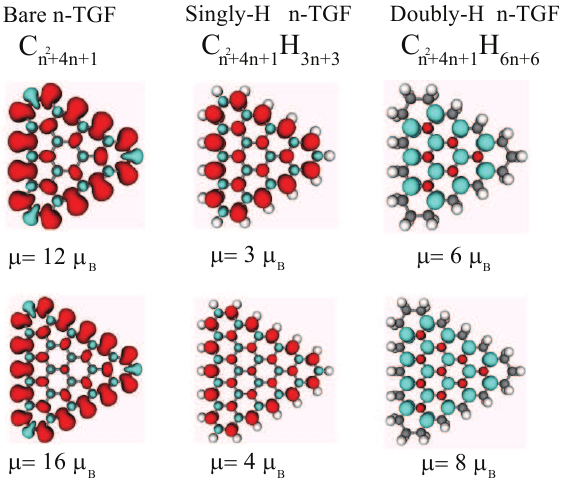}
\caption{(Color online) Atomic structure and isosurface of charge
density difference of spin-up ($\uparrow$) and spin-down ($\downarrow$) states for 4-(upper row) and 5-triangle (lower row) graphene flakes: Bare, singly- and doubly-hydrogenated edges. Calculated net magnetic moments of the flakes are given in terms of Bohr magneton ($\mu_{B}$). Difference charge density of spin-up and spin-down states is shown by red (dark) and blue (light) isosurface, respectively.}\label{charge}
\end{figure}

\section{CALCULATION METHODS}

Optimization of geometrical structures of triangular graphene
flakes and calculations of their magnetic and electric properties
are  performed by using  the  software package Atomistix ToolKit
(ATK) \cite{atk} based on density functional theory (DFT). The
spin-dependent exchange-correlation potential is approximated
within the generalized gradient approximation. The criteria of
convergence used for total energy and Hellman-Feynman forces were
$10^{-4}$ eV and 0.005 eV/\AA, respectively. The electrostatic
potentials were determined on a real-space grid with a mesh cutoff
energy of 150 Ry. Double-zeta-polarized basis sets of local numerical
orbitals were employed to increase the accuracy of our calculations.

For determination of quantum transport properties of the
electrode-TGF-electrode system, ATK use nonequilibrium Green's
function formalism. Transport calculations are performed with the
Brillouin zone sampled with (1,1,51) points within the
Monkhorst-Pack k-point sampling scheme. The current through the
TGFs is determined by summing the transmission probabilities for
electron states from one electrode to another within the energy
window $\mu_{L}-\mu_{R}=|eV|$, where $\mu_{L}$ ($\mu_{R}$) is the
electrochemical potential of the left (right) electrode under the
applied bias $V$. Therefore the spin dependent current is given by
the formula

\begin{equation}
I_\sigma(V)=G_{0}\int^{\mu_{R}}_{\mu_{L}}T_\sigma(E,V)dE
\end{equation}
where $G_{0}=e^{2}/h$ is the quantum conductance unit and
$T_\sigma(E,V)$ is the quantum mechanical transmission probablity
for electrons with spin state $\sigma$ and energy $E$. During the
self-consistent calculation of $I$-$V$ spectrum charge on the TGFs
is not fixed and energy is minimized with respect to the
electrochemical potentials of the electrodes at each voltage
increment. In order to achieve convergence of the electronic
states with increasing voltages to the desired level of accuracy,
calculations performed within the bias window (from -1 to +1 V) in
steps of 0.005 V. By using the carbon chains attached to the TGF,
it is ensured that the screening occurs in the device region.

\begin{figure}
\includegraphics[width=7cm]{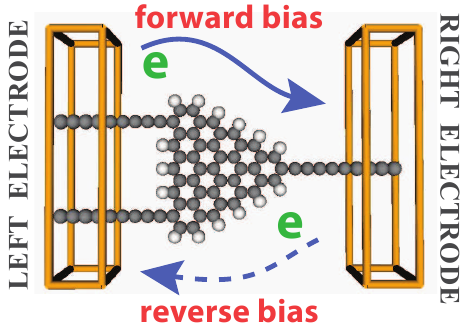}
\caption{(Color online) Electrode-device geometry and
convention for forward and reverse bias applied to triangular graphene
flakes (TGFs).}

\label{iv1}
\end{figure}

\section{Results}

\subsection{Atomic and Magnetic Ground State Properties of TGFs}

First, we have performed geometry optimizations of TGFs and
determined the spin polarized charge density of the optimized
structures. Due to the particular shape of the TGFs spin-relaxed
calculation leads to lower ground state energies compared to those
of spin-unrelaxed calculation; here we have found that
ferromagnetically ordered spin accumulation at the edges gives a
nonzero magnetic moment to the flake. Apart from minor bond
contractions of the edge atoms TGFs preserve their regular
hexagonal structure even when the flakes are not hydrogenated.
Saturating the edge carbon atoms with either one or two hydrogen
atoms considerably modifies the magnetic ground state, by altering
the total magnetic moment.

\begin{figure*}
\includegraphics[width=18cm]{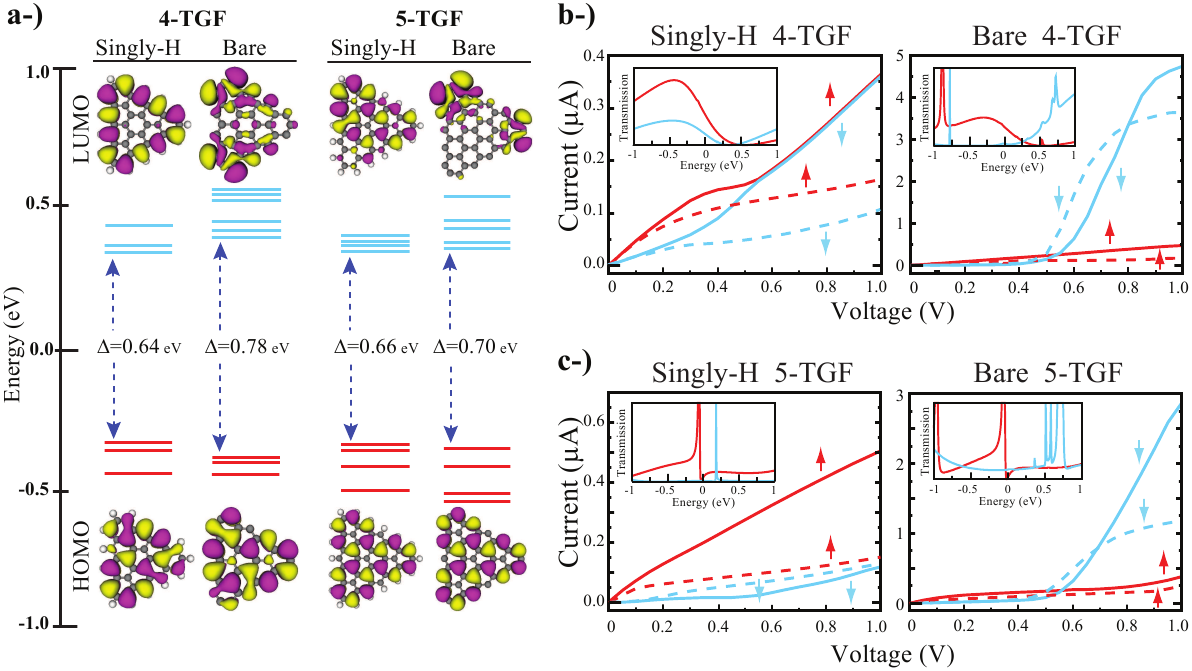}
\caption{(Color online) (a) Energy level spectra within $\pm1$
eV range of Fermi level ($E_{F}$), the HOMO-LUMO gap ($\Delta$) and isosurface of HOMO and LUMO orbitals. (b-c) Calculated I-V curves for hydrogenated and bare triangular graphene flakes (TGFs). Results of 4-TGF and 5-TGF are presented, respectively. Spin-up ($\uparrow$) and spin-down
($\downarrow$) currents are shown by red (dark) and blue
(light) lines, respectively. Solid and dashed lines denote forward
and reverse bias calculations, respectively. Transmission curves
of spin-up and spin-down under zero bias are also shown by
insets. Transmission spectra of singly-H and bare TGFs are plotted
up to maximum value of 0.03 and 0.4, respectively. Fermi levels are set to zero.}
\label{iv2}
\end{figure*}

Graphene, with its hexagonal lattice structure resulting from
$sp^{2}$-type hybridization of carbon atoms, is a planar $\pi$
conjugated system. It can be
viewed as made up of A- and B-sublattices of carbon atoms.
Repulsive case of the Lieb's theorem \cite{lieb} reveals
uniqueness of the ground state and provides a simple formula for
calculating the magnetic ground state of such bipartite systems.
According to the rule provided by the theorem, total net spin
magnetization of a graphene structure is given by

\begin{equation}
\mu=\frac{1}{2}|N_{A}-N_{B}|g \mu_B
\end{equation}

where $g\approx 2$ for electron, $\mu_B$ is Bohr magneton, and $N_{A}$
and
$N_{B}$ denote the number of carbon atoms in A- and B-sublattice,
respectively. Each carbon atom in graphene is connected to the
nearest neighbors by three covalent bonds, while leaving behind a
$p_{z}$-orbital electron contributing to the spin magnetic moment.
Such electrons in each sublattice have opposite spin states and make
spin paired $\pi$-bonds, thus
total spin magnetic moment of the system is zero unless a
difference is created in the numbers of atoms of the A- and
B-sublattices. In some graphene flakes, such as those having
equilateral triangular shapes and zigzag edges, $N_A$ and $N_B$
are different and leads to finite net spin magnetic moments
\cite{kaxiras,chemphys}. Moreover, in bare (not hydrogenated)
flakes the edge carbon atoms that make only two covalent bonds
contribute to the spin moment with two nonbonding electrons.

\begin{figure*}
\includegraphics[width=12cm]{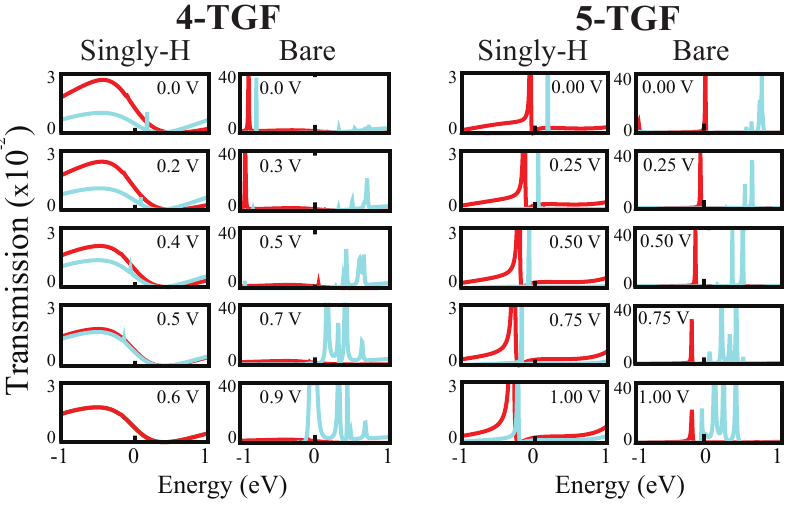}
\caption{(Color online) Bias dependent transmission spectra of
singly hydrogen passivated (singly-H) and bare TGFs for forward bias. Spin-up and spin-down transmissions are shown by red (dark) and blue (light) lines. $E_{F}$ is set to zero.}\label{trans}
\end{figure*}

In Fig.\ref{charge}, we present the optimized structures and
spin charge density difference ($\Delta \rho=\rho_\uparrow -
\rho_\downarrow$) isosurface for bare, singly- and
doubly-hydrogenated cases of $n$-TGFs ($n$=4, 5). In general, we
see that each carbon atom has an induced spin imbalance opposite
to its nearest neighbors, however the spin polarization of hydrogen
atoms is negligible.

Calculated total spin magnetic moments of the structures using DFT
are in integer multiples of $\mu_B$ and either verifies the
results of Lieb's theorem or can be understood by simple
modifications of it. In the flakes with singly-hydrogenated edges,
all the carbon atoms are coordinated as reminiscent of infinite
graphene, and $\mu/\mu_B=N_A-N_B=n-1$ in accordance with Lieb's
theorem. When the hydrogen atoms are removed from the flake (bare
flake case) the $3n$ atoms of sublattice A and 3 atoms of
sublattice B each have an extra nonbonding electron that
contributes to the magnetic moment, consequently giving
$\mu/\mu_B=n-1+3n-3=4(n-1)$. On the other hand, in
doubly-hydrogenated flakes, all the edge carbon atoms have $sp^3$
hybridized electrons with vanishing contribution to spin magnetic
moment. Thus, in this case $\mu/\mu_B=|n-1-3n+3|=2(n-1)$, where
the direction of magnetic moment is reversed, and the net moment
of the system increases with respect to the singly-hydrogenated
case. Tunability of the spin magnetic moments of triangular
graphene flakes through degree of hydrogenation is an interesting
feature that may be utilized for nanodevice applications.
Saturation of TGF edges by other atoms may give rise to similar
modifications in their electronic and magnetic structures.

\subsection{Transport Properties of TGFs}

Next we consider electrical conductance and I-V
characteristics of TGFs when contacted with metallic electrodes.  We
use linear carbon chains as model electrodes.  Carbon atomic
chains which are known to be metallic\cite{tongay,
hasan-transport} are expected to make reasonably good contacts
with the flakes. As an alternative to carbon-chain electrodes, earlier the
robustness of our conductance calculations was also tested by
using semi-infinite gold bar electrodes and consistent results
were obtained for rectangular flakes.\cite{hasan-transport}
At the contact sites the introduced hydrogen atoms are removed and
connection between electrode and flake carbon atoms is established
through double bonds. Due to asymmetric form of device-electrode
geometry one expects different current profiles for forward and
reverse bias voltages. The convention we have adopted is
schematically shown in Fig.\ref{iv1}, where forward bias
corresponds to flow of electrons from the left to the right electrode, i.e. the current is flowing from right to left electrode.

For the ballistic current of electrons, the spin dependent
transmission probability from one electrode to other strongly
depends on the eigenstates of the TGF molecule. The molecular
orbitals perfectly coupled to the electrodes behave as an open
channel and provides propagation with minimal scattering through
the TGF and hence the magnitude of the transmission coefficient is
determined by electrode-device interaction. Eigenstates within the
$E_{F}\pm$1 eV energy window, indicating the
highest-occupied-molecular-orbital (HOMO) and
lowest-unoccupied-molecular-orbital (LUMO) of singly-hydrogenated
and bare n-TGFs (n=4,5) are presented in Fig.\ref{iv2} (a). It is
obvious that the HOMO-LUMO gap ($\Delta$) gets narrower upon the
hydrogenation of TGFs. Resulting from the ferromagnetic ground
state, within the $E_{F}\pm$1 eV energy window up and down spin states are well-separated around the E$_{F}$. In the energy window used for Fig.\ref{iv2} (a), all the filled
levels are for up-spin states, whereas the unoccupied levels
are for down spins. When the TGF molecule is connected to electrodes, however, due to the chain-TGF interactions energy level spectra is changed and it is not easy to exactly distinguish the contribution of TGF and electrode states.

In Fig.\ref{iv2} (b-c), we show the I-V characteristics of 4- and
5-TGFs for bare and singly-hydrogenated cases. Hydrogenated flakes
have lower conductance and leads to at least an order of magnitude
smaller currents, since the hydrogenation of the flake removes
some of the states providing open channels in the flake. While the
maximum current in the calculated bias range is $\sim$ 5.00 $\mu$A
for bare TGF, after hydrogenation it is reduced to $\sim$0.35
$\mu$A.  Since electrode device coupling broadens the
energy levels, and may shift them due to charging, even though
there is no molecular state at E$_{F}$, the tails of HOMO and
LUMO states can contribute to the transmission even at small
voltages. Orbital character that changes upon the H termination of HOMO and LUMO states reveal the importance of the edge atoms in electron transport. In addition, there is a strong rectification of current 
for both singly-H and bare edge cases. In this electrode-device configuration forward 
current gets larger at a threshold bias of $\sim$0.6 V. For small 
voltages, the characteristics of the current flowing through the 
TGFs can be understood by zero bias transmissions that are given 
by insets in Fig.\ref{iv2} (b-c). In fact the self consistently calculated voltage dependent transmission spectra provides a better information regarding the I-V characteristics of the device.

Depending on the applied voltage, triangular graphene flakes
display diverse properties. In the case of forward bias
application to the singly-hydrogenated 4-TGF, the spin-up
current is dominant up to 0.55 V, but after this critical value, the  spin-up and spin-down states are merged due to the transition from ferromagnetic to antiferromagnetic state and thus spin polarizing property disappears. This behavior of I-V curves can be revealed from transmission spectrum. Even if the zero bias
transmissions shown in insets in Fig.\ref{iv2} (b-c) can explain I-V curves for small bias values, at finite bias voltages the
transmission spectrum of the flake changes through broadening
and/or shifts of the transmission peaks, which should be
calculated self-consistently under nonequilibrium conditions. In
Fig.\ref{trans} we show the variations of transmission spectrum for
selected cases under incremental forward bias voltages. For the
hydrogenated 4-TGF, while transmission of the up spin
channel decreases with increasing voltages, the transmission of
down spins is increasing. Eventually both transmission curves
are merging at the vicinity of 0.55 V and hence the spin
polarization of the current is ceasing. Total magnetic moment of the flake together with electrodes has a bias dependence, gradually
decreasing from 2 $\mu_{B}$ to zero at the merging point of the
up and down spin currents. In contrast, the down spin
states of bare 4-TGF under forward bias get closer to the E$_{F}$ and become dominant carriers in the current. This explains
how the spin polarization of the current is switched by applied
bias. We can also establish a relation between I-V behavior and
the corresponding transmission spectrum of bare 5-TGF. Since
up and down spin transmission  peaks do not show considerable variation within the energy window, the spin polarization
of the currents is maintained up to 1 eV.

\section{CONCLUDING REMARKS}

In summary, we have investigated the electric, magnetic and
transport properties of triangle shaped graphene flakes. We have found
that in addition to the their ferromagnetic ground state, triangular graphene flakes show spin polarized and rectified current properties depending on edge saturation, flake size, bias voltage and bias direction. Diverse and spin dependent properties of graphene flakes depending on their shape, size and edge saturation keep the promise of variety of application in future nanospintronics.

\section{ACKNOWLEDGMENTS}

This work was supported by T\"{U}B\.{I}TAK under Grant No.
106T597, and through TR-Grid e-Infrastructure Project. Computing
resources used in this work were partly provided by the National Center
for High Performance Computing of Turkey (UYBHM) under Grant No.
2-024-2007.


\begin{thebibliography}{99}


\bibitem{novo1} K. S. Novoselov, A. K. Geim, S. V. Morozov, D. Jiang,
Y. Zhang, S. V. Dubonos, I. V. Grigorieva, and A. A. Firsov,
 Science \textbf{306}, 666 (2004).

\bibitem{novo2} K. S. Novoselov, A. K. Geim, S. V. Morozov, D. Jiang,
M. I. Katsnelson, I. V. Grigorieva, S. V. Dubonos, and A. A. Firsov,
 Nature \textbf{438}, 197 (2005).

\bibitem{zhang} Y. Zhang, Y.-W. Tan, H. L. Stormer, and Philip Kim,
 Nature (London) \textbf{438}, 201 (2005).

\bibitem{morozov}
S. V. Morozov, K. S. Novoselov, M. I. Katsnelson, F. Schedin, D. C.
Elias, J. A. Jaszczak, and A. K. Geim, Phys. Rev. Lett. 100, 016602
(2008).

\bibitem{lee}
C. Lee, X. Wei, J. W. Kysar, and J. Hone, Science 321, 385 (2008).

\bibitem{bolotin} K. I. Bolotin, F. Ghahari, M. D. Shulman, H. L.
Stormer,  and  P. Kim,
 Nature (London) \textbf{462}, 196 (2009).

\bibitem{ultrafast} F. Xia, T. Mueller, Yu-ming Lin, A. Valdes-Garcia,
and P. Avouris,
 Nature Nanotechnology \textbf{4}, 839 (2009).

\bibitem{readout} C. Chen, S. Rosenblatt, K. I. Bolotin, W. Kalb, P.
Kim, I. Kymissis, H. L. Stormer, T. F. Heinz, and  J. Hone,
 Nature Nanotechnology \textbf{4}, 861 (2009).

\bibitem{wang}
X. Wang, L. Zhi, and K. Mullen, Nano Lett. 8, 323 (2008).
\bibitem{kim}
K. S. Kim, Y. Zhao, H. Jang, S. Y. Lee, J. M. Kim, K. S. Kim, J.-H. Ahn,
P. Kim, J.-Y. Choi, and B. H. Hong, Nature (London) 457,706 (2009).



\bibitem{yan}
Q. Yan, B. Huang, J. Yu, F. Zheng, J. Zang, J. Wu, B.-L. Gu, F. Liu, and
W. Duan, Nano Lett. 7, 1469 (2007).

\bibitem{murali}
R. Murali, K. Brenner, Y. Yang, T. Beck, and J. D. Meindl, IEEE Electron
Device Lett., 30,611 (2009).

\bibitem{edge-states} H. Lee, Y. Son, N. Park, S. Han, and J. Yu,
 Phys. Rev. B \textbf{72}, 174431 (2005).

\bibitem{cohen} Y. Son, M. L. Cohen, and S. G. Louie,
 Nature (London) \textbf{444}, 347 (2006).

\bibitem{hasan-transport} H. \c{S}ahin and R. T. Senger,
 Phys. Rev. B \textbf{78}, 205423 (2008).

\bibitem{ezawa2} M. Ezawa,
 Phys. Rev. B \textbf{76}, 245415 (2007).

\bibitem{hod2} O. Hod, J. E. Peralta, and G. E. Scuseria,
 Phys. Rev. B \textbf{76}, 233401 (2007).

\bibitem{defect1} O. V. Yazyev and L. Helm,
 Phys. Rev. B \textbf{75}, 125408 (2007).

\bibitem{defect2} J. J. Palacios, J. Fernandez-Rossier, and L. Brey,
 Phys. Rev. B 77, 195428 (2008).

\bibitem{hod} O. Hod, V. Barone, and G. E. Scuseria, Phys. Rev. B 77,
035411 (2008).



\bibitem{akola} J. Akola, H. P. Heiskanen, and M. Manninen,
 Phys. Rev. B \textbf{77}, 193410 (2008).

\bibitem{kaxiras} W. L. Wang, S. Meng, and E. Kaxiras
 Nano Letters \textbf{8}, 241 (2008).

\bibitem{chemphys} M. R. Philpott, F. Cimpoesu, and Y. Kawazoe,
 Chem. Phys. \textbf{354}, 1 (2008).

\bibitem{Rozhkov} A. V. Rozhkov and Franco Nori, Phys. Rev. B
\textbf{81}, 155401 (2010)

\bibitem{hasan-apl} H. \c{S}ahin, C. Ataca, and S. Ciraci,
 Appl. Phys. Lett. \textbf{95}, 222510 (2009).


\bibitem{elias} D. C. Elias, R. R. Nair, T. M. G. Mohiuddin, S. V.
Morozov, P. Blake, M. P. Halsall, A. C. Ferrari, D. W. Boukhvalov, M.
I. Katsnelson, A. K. Geim and K. S. Novoselov, Science \textbf{323},
610 (2009).

\bibitem{sofo} J. O. Sofo, A. S. Chaudhari and G. D. Barber,
Phys. Rev. B \textbf{75}, 153401 (2007).

\bibitem{boukhvalov} D. W. Boukhvalov, M. I. Katsnelson and
A. I. Lichtenstein, Phys. Rev. B \textbf{77}, 035427 (2008).


\bibitem{Graphene to graphane} M. Z. S. Flores, P. A. S. Autreto, S. B.
Legoas and D. S. Galvao , Nanotechnology \textbf{20}, 465704 (2009).


\bibitem{graphaneNR} H. \c{S}ahin, C. Ataca, and S. Ciraci,
 Phys. Rev. B \textbf{81}, 205417 (2010).

\bibitem{Bekyarova} E Bekyarova, M. E. Itkis, P. Ramesh, C. Berger, M. Sprinkle, W. A. de Heer and R. C. Haddon, J. Am. Chem. Soc. \textbf{131} 1336 (2009).

\bibitem{Boukhvalov} D. W. Boukhvalov and M. I. Katsnelson Phys, Rev. B
\textbf{78} 085413 (2008).

\bibitem{Pei} Qing-Xiang Pei, Yong-Wei Zhang, V. B Shenoy
Nanotechnology \textbf{21}, 115709 (2010)


\bibitem{Warner} J. H. Warner, M. H. Rummeli, Ling Ge, T. Gemming, B.
Montanari, N. M. Harrison, B. Buchner, G. Andrew D. Briggs, Nature
Nanotechnology \textbf{4}, 500 (2009).

\bibitem{girit} J. C.Meyer, C. O. Girit, M. F. Crommie and A. Zettl,
Nature \textbf{454}, 319 (2008).


\bibitem{Dresselhaus} X. Jia, M. Hofmann, V. Meunier, B. G. Sumpter, J.
Campos-Delgado, José M. Romo-Herrera, H. Son, Ya-Ping Hsieh, A. Reina,
J. Kong, M. Terrones, M. S. Dresselhaus, Science \textbf{323}, 1701
(2009).


\bibitem{dobrik} L. Tapaszto, G. Dobrik, P. Lambin, L. P. Biro, Nat.
Nanotechnol. \textbf{3}, 397 (2008).

\bibitem{ozyilmaz} M. Y. Han, B. Ozyilmaz, Y. Zhang, P. Kim, Phys. Rev.
Lett. \textbf{98}, 206805 (2007).


\bibitem{gan} C. K. Gan and D. J. Srolovitz, Phys. Rev. B 81, 125445
(2010)

\bibitem{olcay} O. U. Akturk and M. Tomak, Appl. Phys. Lett. \textbf{96}, 081914 (2010).

\bibitem{wei} W. Zhang, L. Sun, Z. Xu, A. V. Krasheninnikov, P. Huai,
Z. Zhu and F. Banhart, Phys. Rev. B \textbf{81}, 125425 (2010).

\bibitem{hakinen} P. Koskinen, S. Malola, and H. H\"{a}kkinen, Phys. Rev. B
\textbf{80}, 073401 (2009).

\bibitem{kuc} A. Kuc, T. Heine, and G. Seifert, Phys. Rev. B
\textbf{81}, 085430 (2010)

\bibitem{atk} Distributed by QuantumWise company, Copenhagen,
 Denmark. http://www.quantumwise.com

\bibitem{lieb} E. H. Lieb,
 Phys. Rev. Lett. \textbf{62}, 1201 (1989).


\bibitem{tongay} S. Tongay, R. T. Senger, S. Dag, and S. Ciraci,
 Phys. Rev. Lett. \textbf{93}, 136404 (2004).

\end{thebibliography}
\end{document}